\documentclass{article}
\usepackage{spconf,amsmath,graphicx}
\usepackage{url}
\usepackage{url}
\usepackage{indentfirst}
\usepackage[draft]{hyperref}
\usepackage{multirow}
\usepackage{amsmath} 
\usepackage{hhline}
\usepackage[export]{adjustbox}

\title{Patient-Specific Finetuning of Deep Learning Models for Adaptive Radiotherapy in Prostate CT}
\name{Mohamed S. Elmahdy$^{\dagger\star}$,  Tanuj  Ahuja$^{\dagger\circ\star}$, U. A. van der Heide$^{\ddagger\bullet}$, and Marius Staring$^{\dagger\ddagger}$\thanks{$^{\star}$denotes equal contribution}}
\address{$^{\dagger}$Department of Radiology, Leiden University Medical Center, Leiden, The Netherlands. \\ $^\circ$ Computer Science and Engineering, Guru Gobind Singh Indraprastha University, Delhi, India. \\ $^{\ddagger}$ Department of Radiation Oncology, Leiden University Medical Center, Leiden, the Netherlands. \\
	$^{\bullet}$ Department of Radiation Oncology, the Netherlands Cancer Institute, Amsterdam, the Netherlands.}

\begin{document}
	%
	\maketitle
	\begin{abstract}
		Contouring of the target volume and Organs-At-Risk (OARs) is a crucial step in radiotherapy treatment planning. In an adaptive radiotherapy setting, updated contours need to be generated based on daily imaging. In this work, we leverage personalized anatomical knowledge accumulated over the treatment sessions, to improve the segmentation accuracy of a pre-trained Convolution Neural Network (CNN), for a specific patient. We investigate a transfer learning approach, fine-tuning the baseline CNN model to a specific patient, based on imaging acquired in earlier treatment fractions. The baseline CNN model is trained on a prostate CT dataset from one hospital of 379 patients. This model is then fine-tuned and tested on an independent dataset of another hospital of 18 patients, each having 7 to 10 daily CT scans. For the prostate, seminal vesicles, bladder and rectum, the model fine-tuned on each specific patient achieved a Mean Surface Distance (MSD) of $1.64 \pm 0.43$ mm, $2.38 \pm 2.76$ mm, $2.30 \pm 0.96$ mm, and $1.24 \pm 0.89$ mm, respectively, which was significantly better than the baseline model. The proposed personalized model adaptation is therefore very promising for clinical implementation in the context of adaptive radiotherapy of prostate cancer. 
	\end{abstract}
	\section{Introduction}
	\label{sec:intro}
	
	\noindent Prostate cancer is the second leading cause of cancer death in American men, behind lung cancer. About 1 man in 41 will die of prostate cancer. The American Cancer Society’s estimates about 174,650 new cases of prostate cancer and about 31,620 deaths from prostate cancer in the United States for 2019 only \cite{NCS}. Radiation Therapy (RT) is one of the treatment options for prostate cancer that uses a highly localized dose distribution to kill cancer cells. RT dose is usually fragmented over 4 to 8 weeks resulting in 20 to 40 daily fractions \cite{Zhang}. Since the dose is delivered in several sessions, variations in the size and shape of the target area and Organs-At-Risk (OARs) is bound to take place. Often these changes are due to organ deformation such as variations in the rectum and bladder filling \cite{Sonke}. Continuing RT using the initial planning CT scan and the corresponding dose distribution despite these changes may lead to under-dosing of the target area or over-dosing the OARs. \\
	\indent Adaptive radiation therapy aims to adapt the treatment dose distribution to the daily anatomy, in order to achieve safe dose escalation \cite{Yan} or smaller margins. This adaptation can be done by re-imaging, re-contouring, and re-planning at every session. However, manual re-contouring takes a lot of time and consequently new variations in the size and the shape of the OARs could be introduced meanwhile. Therefore, there is a need to re-contour the CT images accurately and in a time-efficient manner, i.e. (semi-)automatically. \\
	\indent Literature of prostate segmentation can be broadly categorized into two approaches, non-learning and learning approaches. For non-learning approaches, Qiao \emph{et al.} used the open source \texttt{elastix} software to apply deformable image registration and subsequently propagate the contours from the planning to the daily scans \cite{Qiao}. For learning-based approaches, Elmahdy \emph{et al.} proposed a hybrid learning and iterative approach, where they used a CNN network to segment the bladder and explicitly feed it to the registration model as prior knowledge on the underlying anatomy \cite{Elmahdy}. A substantial improvement was observed compared to the results reported in \cite{Qiao}. Recently, deep learning and specifically convolutional neural networks (CNNs) are being used to automatically segment OARs. Deep learning has the power to extract information from the data rather than depending on hand-crafted features. Milletari \emph{et al.} developed an end-to-end CNN network to segment 3D medical images and reported a Dice Similarity Coefficient (DSC) of 0.869 $\pm$ 0.033 on the PROMISE 2012 challenge dataset \cite{Milletari} (MR data). Tong \emph{et al.} proposed the use of a shape representation model, to learn highly representative shape characteristics of OARs and help the final segmentation of a Fully Convolution Neural Network (FCNN) \cite{Tong}. 
	
	To the best of our knowledge, no deep learning based approach has been used to adapt a CNN model to a particular patient at a certain time point. In this study, we propose to adapt a baseline CNN model as the patient goes through their RT treatment. Thus, instead of depending on a static deep learning model, we accumulate the knowledge over successive sessions for the same patient. This accumulated knowledge is then used to encourage the model into predicting a more personalized segmentation. Moreover, we study the performance of the fine-tuned network when more and more imaging becomes available over the course of treatment.
	
	\section{MATERIALS AND METHODS}
	\subsection{Dataset}\label{sec:materials}
	
	\noindent This study includes two different datasets from two different institutes and scanners, for patients who underwent intensity-modulated RT for prostate cancer. First, a dataset from Leiden University Medical Center (LUMC), Netherlands, has a total of 379 patients with one CT scan each. The scans were acquired using a Toshiba scanner, having 68 to 240 slices with a voxel size of approximately 1.0 $\times$ 1.0 $\times$ 3.0 mm. The second dataset is from Haukeland University Hospital, Norway, and includes 18 patients with 8-11 CT scans each corresponding to multiple fractions. These scans were acquired using a GE scanner, having 90 to 180 slices with a voxel size of approximately 0.9 $\times$ 0.9 $\times$ 3.0 mm. The target structures (prostate and seminal vesicles) as well as OARs (bladder and rectum) were manually delineated by oncologists. For the LUMC data informed consent was waived by the local Medical Ethical Committee, while for the Haukeland data informed consent was given by all included patients. 
	
	\subsection{Baseline Segmentation CNN Model}
	\noindent In this study, for the baseline CNN model we adopted the network introduced in Elmahdy \emph{et al.} \cite{Elmahdy}, which has a straightforward architecture, but increased the number of the output labels from 2 to 5 for multi-organ segmentation. Similar to the standard U-net, it has an encoder and decoder path with four resolutions. Each encoder block consists of 3 $\times$ 3 $\times$ 3 convolutions with a stride of two in each dimension, followed by a rectified linear unit (ReLu). In the synthesis path, each block of the decoder consists of 3 $\times$ 3 $\times$ 3 convolutions with a stride of one in each dimension, followed by a ReLu and an upsampling of 2 $\times$ 2 $\times$ 2. For a better localization, high resolution features from the contracting path are combined with the upsampled output, using skip connections. This aids to recover fine-grained details that were lost during the compression phase. The upsampling path is then followed by one Fully Connected (FC) layer, similar to the original paper, or three FC layers. These models are denoted by \textbf{base$^a$} and \textbf{base$^b$}, respectively. A softmax layer finalizes the models.
	
	\subsection{Patient-specific Daily Model Adaptation}
	\noindent In order to adapt the model to the patient-specific anatomy, we update the model with the anatomical knowledge from previous treatment sessions for the same patient. We thus personalize the pre-trained baseline CNN and utilize the available imaging of the patient. Since the new data (daily scans) as well as the task (segmentation) are the same as the base models, we re-use the base model weights as initialization to the model adaptation, as in a transfer learning approach. The weights in the FC layers are then fine-tuned using the imaging from previous fractions, while all other weights remain fixed. For each patient, at the first fraction we fine-tune the base model based on the planning scan. At subsequent fractions we start with the previous fine-tuned model and continue fine-tuning based on the scan and segmentation pair of the previous fraction. Here we consider that the segmentation can be manually corrected to clinical quality before the start of the current fraction. To be precise, let $M_0$ be the base model and $M_j$ the fine-tuned model at fraction $j$, $\Phi$ the model adaptation process, $(I_j, S_j)$ the image and segmentation pair at fraction $j$. Model adaptation is then performed as follows:
	\begin{align}\label{eq:adaptive}
	M_j = \Phi( M_{j-1}; I_{j-1}, S_{j-1} ).
	\end{align}
	The segmentation prediction of the current fraction is then given by $M_j( I_j )$. The base models are trained on the LUMC dataset. The networks are adapted to the Haukeland dataset, which has multiple follow-up scans per patient, mimicking an adaptive RT scenario. 
	
	\section{Experiments and Results}
	
	\subsection{Evaluation Measures and Implementation} 
	\noindent The Dice Similarity Coefficient (DSC), Mean Surface Distance (MSD), and 95\% Hausdorff Distance (HD) are used to evaluate the error between the ground truth delineations and the segmentations predicted by the models. These are evaluated for the prostate, seminal vesicles, bladder, and rectum. A Wilcoxon signed rank test at $p = 0.05$ is used to assess statistical significance. Evaluation results are shown for the Haukeland data only, as only this dataset has follow-up scans. 
	
	We used the Tensorflow library for the implementation of the 3D CNNs. Out of the 379 LUMC patients, 70\% (259 cases) were used for training the baseline models and the remaining 30\% (111 cases) were used for validation. A total of 1000 patches of size 128 $\times$ 128 $\times$ 128 were extracted from each CT volume, sampling the classes with a uniform distribution to handle class imbalance. All scans were resampled to a fixed voxel size of 1 $\times$ 1 $\times$ 2 mm in order to handle variations in voxel size. The Dice Similarity Coefficient (DSC) is deployed as a cost function and the network was trained using the state-of-the-art Rectified Adam (RAdam) optimizer introduced in \cite{Reddi}, with a fixed learning rate of $10^{-4}$. The network was trained for 1,000,000 iterations with a batch size of 4 using an NVIDIA Titan Xp GPU with 12 GB of memory. For model adaptation, a total of 2000 patches were extracted from a single CT volume with the same patch size, cost function, and learning rate as the base model. For investigating the effect of the number of adaptation iterations on the network performance, we varied this parameter between 500, 2000, and 5000 iterations. This adaptation can be performed offline, and took less than an hour per fraction at 2000 iterations.
	
	\begin{table}[t]
		\centering
		\setlength{\tabcolsep}{3.5pt}
		\caption[Table caption text]{ DSC value of the target volumes and OARs.}
		\resizebox{0.485\textwidth}{!}{
			\begin{tabular}{c*{7}{c}c}
				&$\#$Itr&Prostate$^{\dagger\ddagger}$&SV$^{\dagger\ddagger}$&Rectum$^{\dagger\ddagger}$& Bladder$^{\dagger\ddagger}$ \\ \hline
				&& $\mu \pm \sigma$ & $\mu \pm \sigma$ & $\mu \pm \sigma$ & $\mu \pm \sigma$ \\ \hline
				base$^a$ && $0.76 \pm 0.06$ & $0.55 \pm 0.19$ & $0.79 \pm 0.08$ & $0.86 \pm 0.14$ \\
				
				proposed$^a$ &500& $0.78 \pm 0.06$ & $0.61 \pm 0.17$ & $0.79 \pm 0.07$ & $0.89 \pm 0.09$ \\
				
				proposed$^a$ &2000& $0.80 \pm 0.05$ & $0.63 \pm 0.17$ & $0.80 \pm 0.07$ & $0.90 \pm 0.07$ \\
				
				proposed$^a$ &5000& $0.82 \pm 0.05$ & $\textbf{0.65} \pm \textbf{0.16}$ & $0.80 \pm 0.07$ & $\textbf{0.91} \pm \textbf{0.07}$ \\
				
				\hline
				
				base$^b$&& $0.77 \pm 0.07$ & $0.46 \pm 0.20$ & $0.78 \pm 0.08$ & $0.87 \pm 0.08$  \\ 
				
				proposed$^b$&500& $0.80 \pm 0.05$ & $0.50 \pm 0.18$ & $0.81 \pm 0.06$ & $\textbf{0.91} \pm \textbf{0.07}$  \\ 
				
				proposed$^b$&2000& $\textbf{0.83} \pm \textbf{0.04}$ & $0.56 \pm 0.15$ & $\textbf{0.83} \pm \textbf{0.05}$ & $\textbf{0.91} \pm \textbf{0.06}$  \\
				
				proposed$^b$&5000& $\textbf{0.83} \pm \textbf{0.04}$ & $0.55 \pm 0.17$ & $\textbf{0.83} \pm \textbf{0.05}$ & $\textbf{0.91} \pm \textbf{0.06}$  \\
				
				\hline
				
		\end{tabular}}
		\label{table:dsc}
	\end{table}
	
	\subsection{Results}
	\noindent Tables \ref{table:dsc}, \ref{table:msd} and \ref{table:hd} show the quantitative results of all the models, averaged over all the treatment fractions. Adapted models are denoted by proposed$^a$ and proposed$^b$. Here, $\dagger$ and $\ddagger$ represent a statistically significant difference between base$^a$ vs proposed$^a$, and base$^b$ vs proposed$^b$ at 2000 iterations, respectively. In terms of DSC, MSD, and 95\% HD, the larger base$^b$ model is slightly better than base$^a$, except for the seminal vesicles. This pattern is also visible for the proposed models. All proposed models outperform the baseline models, sometimes by a margin. Furthermore, adapting the network for 5000 iterations seems as good as 2000 iterations as shown in the tables. Predicting a segmentation took $\sim$0.6 seconds.
	
\begin{figure*}[ht]
		\centering
		\resizebox{\textwidth}{!}{
			\begin{tabular}{ c @{\quad} c @{\quad} c @{\quad} c}
				
				\multirow{-15}{*}{\rotatebox[origin=c]{90}{\Huge{base$^b$}}} &
				\includegraphics[width=90mm,height=60mm]{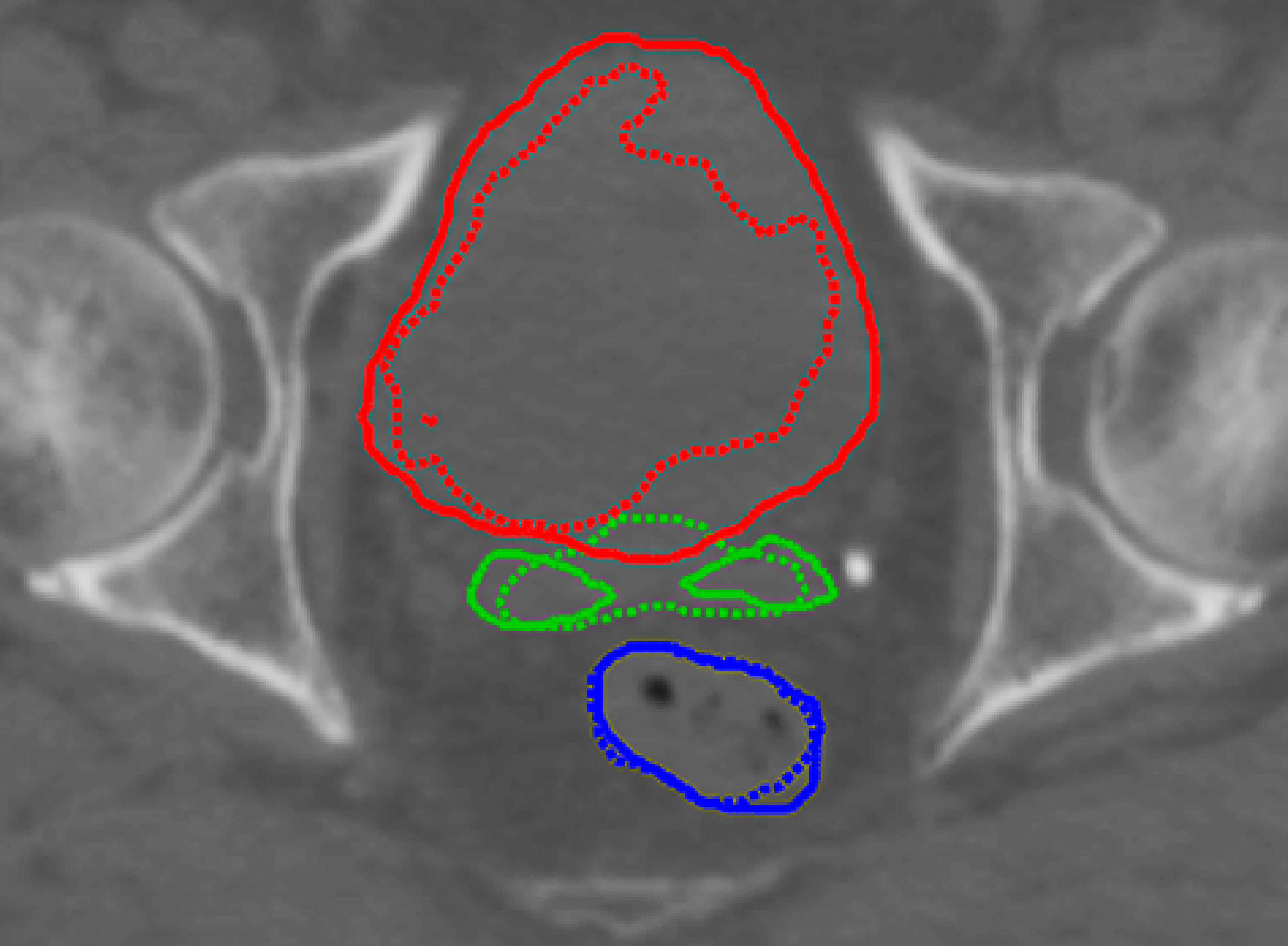} &
				\includegraphics[width=90mm,height=60mm]{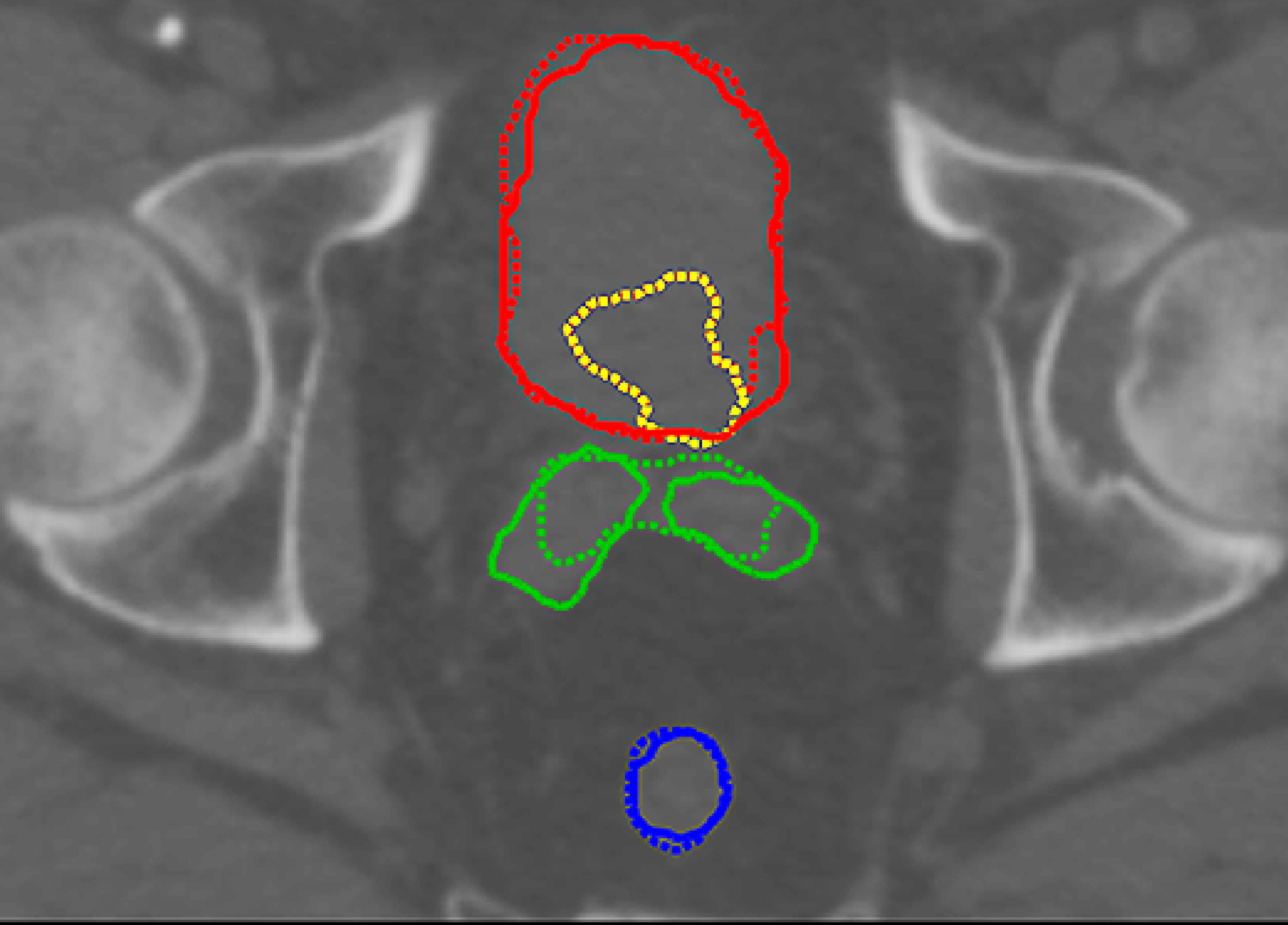} &
				\includegraphics[width=90mm,height=60mm]{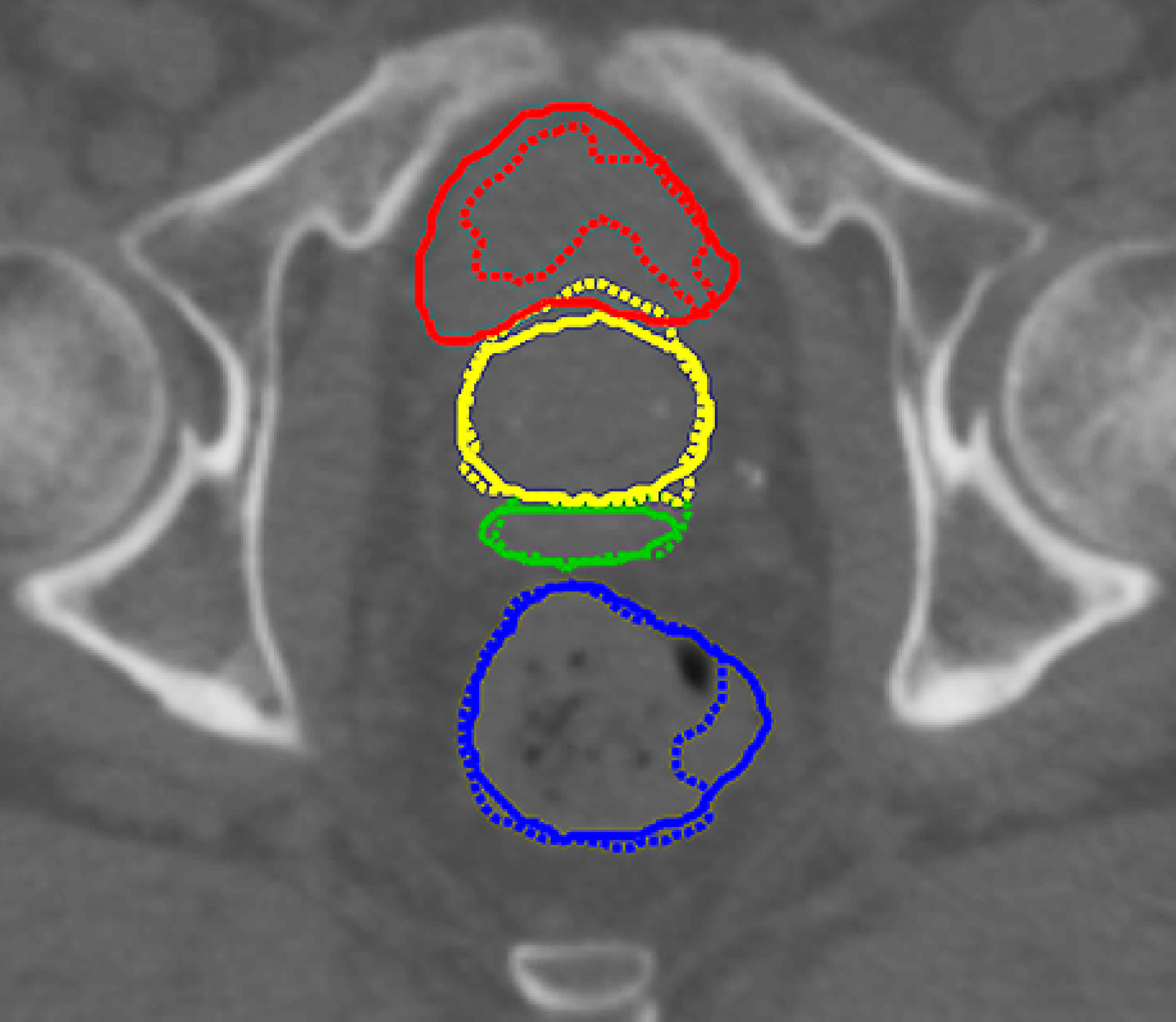} \\ 
				
				\multirow{-15}{*}{\rotatebox[origin=c]{90}{\Huge{proposed$^b$}}} &
				\includegraphics[width=90mm,height=60mm]{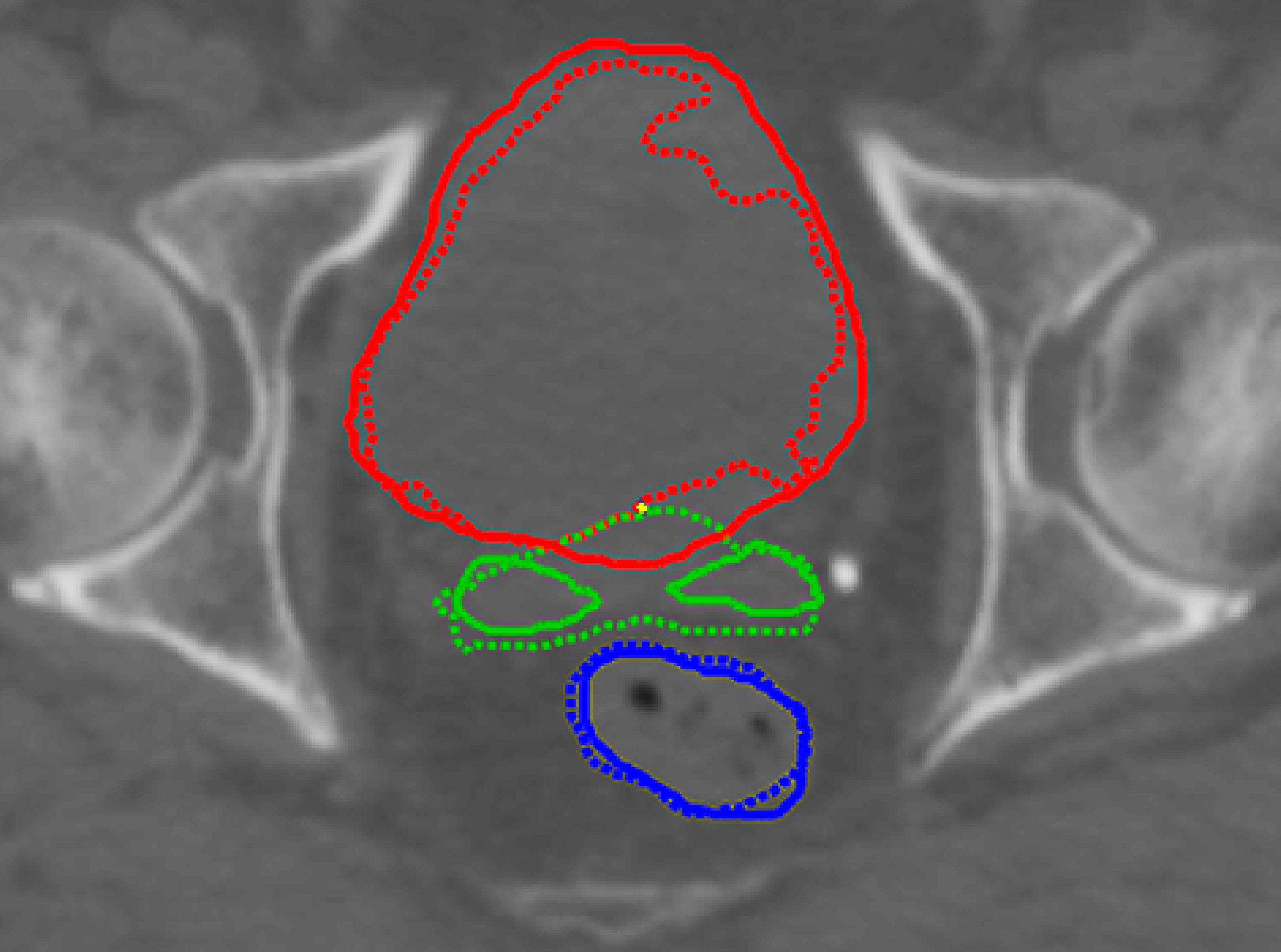} &
				\includegraphics[width=90mm,height=60mm]{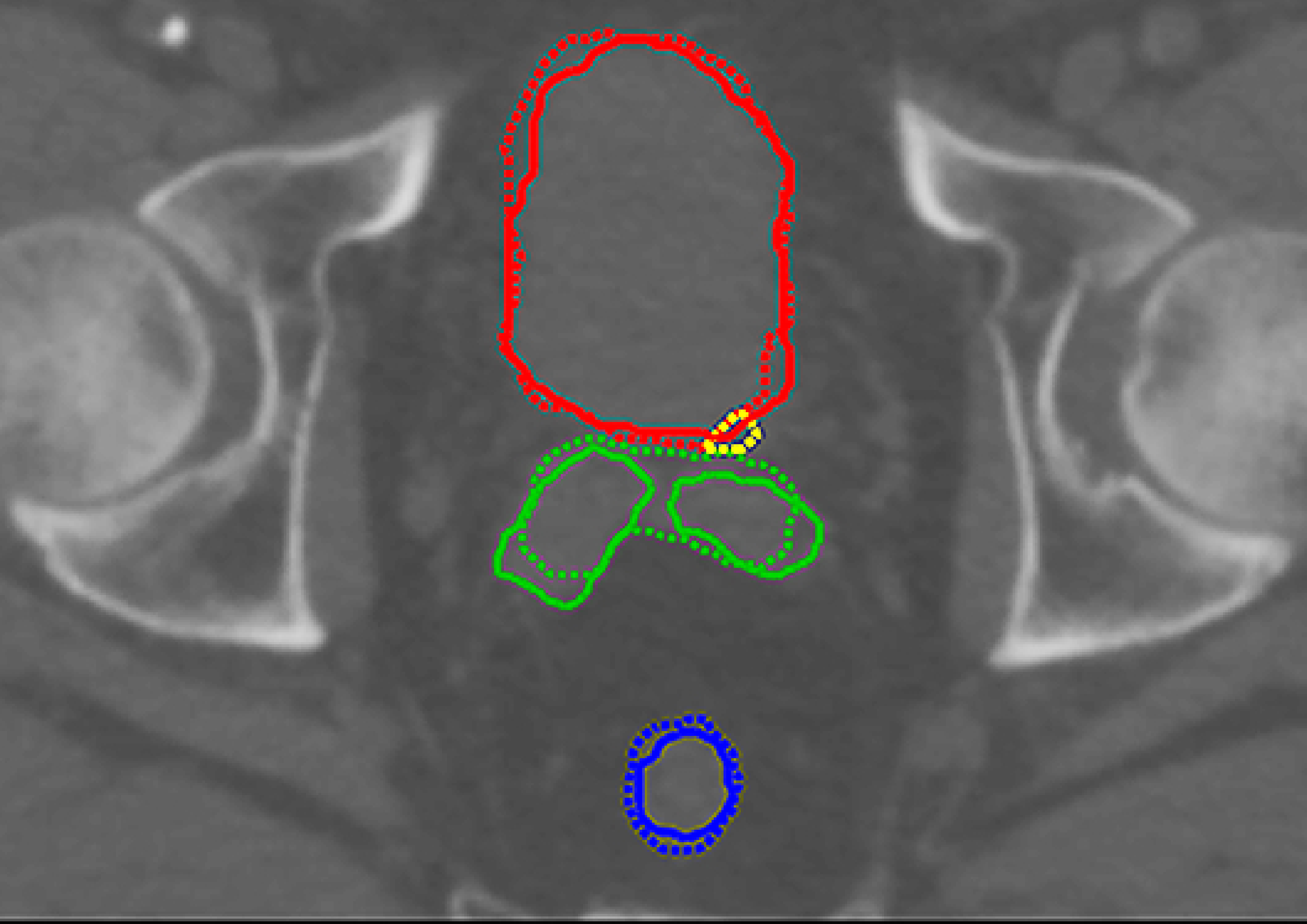} &
				\includegraphics[width=90mm,height=60mm]{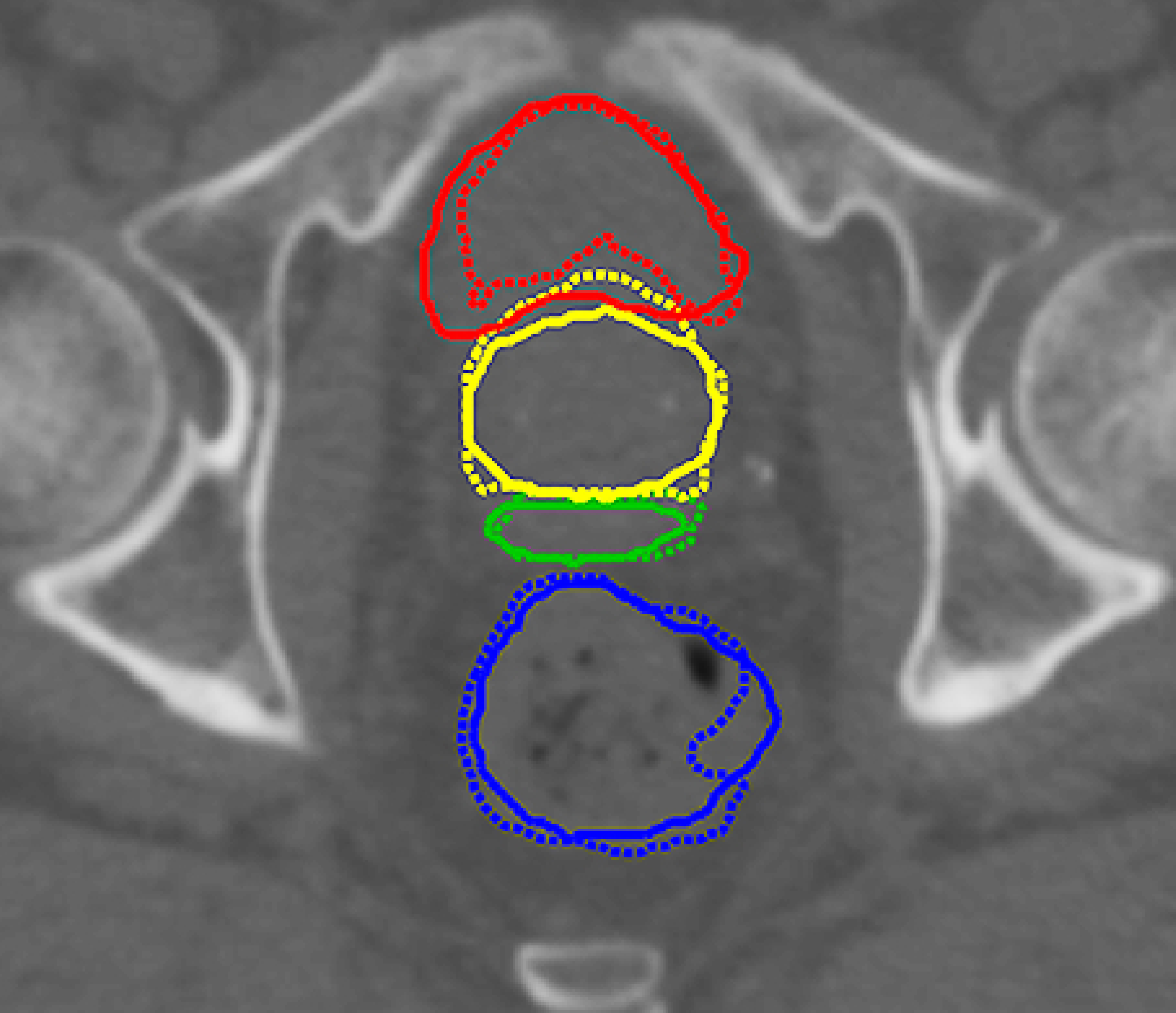} \\ 
				
			\end{tabular}
		}
		\caption{Three example results for the base$^b$ model (top row) and the proposed$^b$ at 2000 itr model (bottom row). The red, yellow, blue and green contours represent the bladder, prostate, rectum, and seminal vesicles, respectively. The solid line indicates manual contouring, while the dotted line the automatic prediction.}
		\label{fig:examples}
	\end{figure*}

	To investigate the relation between the amount of patient-specific data available for fine-tuning and segmentation performance, we applied the models $M_j, j \in \{0,5\}$ on scan $I_6$ from each patient. Boxplots of the results are shown in Figure \ref{fig:msd} for the MSD. Segmentation of most structures tend to improve when more data is available, while such a trend is not visible for the rectum. Some example segmentation results are given in Figure \ref{fig:examples} for the models base$^b$ and proposed$^b$ at 2000 iterations.
	
	\section{Discussion and Conclusion}
\begin{figure*}[ht]
		\centering
		\resizebox{\textwidth}{!}{
			\includegraphics[width=\linewidth, height=53.2mm]{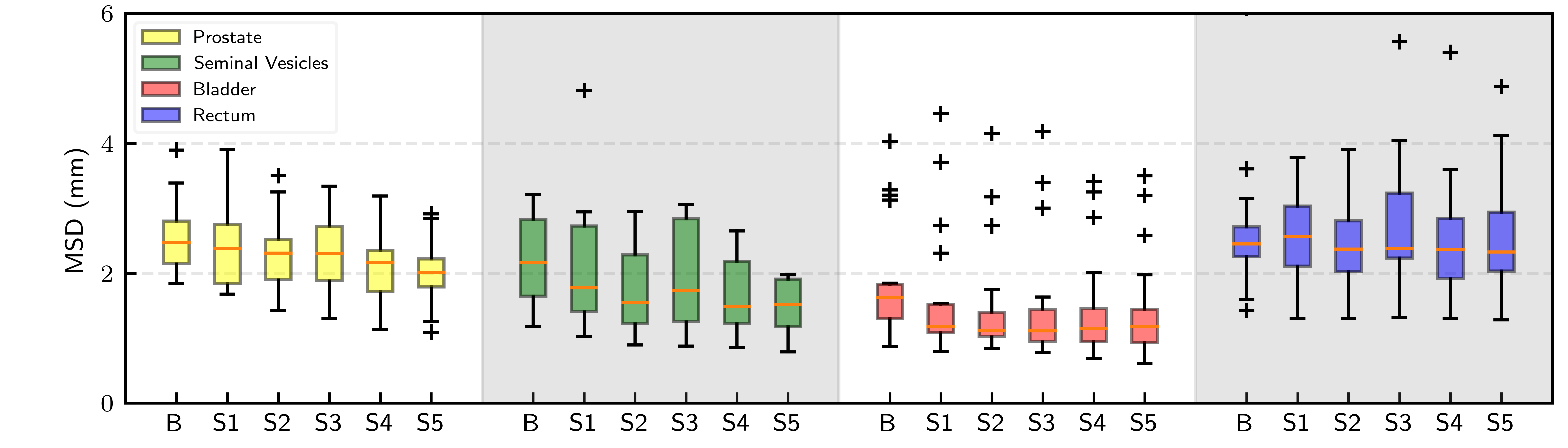}
		}
		\caption{The effect of model adaptation over sessions in terms of MSD (mm). B is base model, S\emph{i} after session $i$.}
		\label{fig:msd}
	\end{figure*}

	\noindent  In this study we investigated the hypothesis of personalizing the automatic contouring by deep learning, of the target organs and OARs for adaptive radiotherapy of prostate cancer. Unlike traditional CNN networks, the proposed model adaptation strategy predicts the segmentation of the daily CT based on the accumulated anatomical knowledge from the previous scans of the same patient.
	We demonstrate that adapting the model to a specific patient anatomy boosted the performance of the network. Furthermore, increasing the network capacity by adding more fully connected layers was beneficial. Moreover, from Fig. \ref{fig:msd} we observe that the rectum delineation did not benefit from fine-tuning, which may be explained by the normal variation in rectum filling over treatment fractions. Such variations in the rectum and seminal vesicles could be further investigated in future work by more sophisticated networks and model generalization methods.
	
	To conclude, we proposed an adaptive training mechanism for personalized automatic contour segmentation for prostate cancer. This adaptation showed potential for improving the prediction of the daily anatomy based on personalized imaging accumulated over factions. Since the segmentation time is less than a second, the adaptation mechanism is therefore very promising for clinical implementation in the context of adaptive radiotherapy of prostate cancer. \\
	
	\noindent \textbf{Acknowledgements.} This study was financially supported by Varian Medical Systems and ZonMw, grant number 104003012. The Haukeland dataset was provided by oncologist Svein Inge and physicist Liv Bolstad; they are gratefully acknowledged.
	
	\begin{table}[t]
		\centering
		\setlength{\tabcolsep}{3.5pt}
		\caption[Table caption text]{ MSD value of the target volumes and OARs.}
		\resizebox{0.485\textwidth}{!}{
			\begin{tabular}{c*{7}{c}c}
				&$\#$Itr&Prostate$^{\dagger\ddagger}$&SV$^{\dagger\ddagger}$&Rectum$^{\dagger\ddagger}$& Bladder$^{\dagger\ddagger}$ \\ \hline
				&& $\mu \pm \sigma$ & $\mu \pm \sigma$ & $\mu \pm \sigma$ & $\mu \pm \sigma$ \\ \hline
				
				base$^a$&& $2.70 \pm 0.88$ & $3.78 \pm 4.76$ & $2.77 \pm 1.23$ & $2.66 \pm 9.46$ \\ 
				
				proposed$^a$ & 500& $2.38 \pm 0.64$ & $2.58 \pm 2.67$ & $2.76 \pm 1.05$ & $1.62 \pm 1.28$  \\ 
				
				proposed$^a$ & 2000& $2.03 \pm 0.55$ & $\textbf{2.38} \pm \textbf{2.76}$ & $2.66 \pm 1.03$ & $1.48 \pm 1.17$  \\
				
				proposed$^a$ & 5000& $1.78 \pm 0.52$ & $2.41 \pm 3.17$ & $2.59 \pm 1.00$ & $1.39 \pm 1.12$  \\ \hline
				
				base$^a$&& $2.29 \pm 0.70$ & $5.08 \pm 5.33$ & $2.78 \pm 1.17$ & $1.83 \pm 0.96$ \\
				
				proposed$^b$  &500& $1.95 \pm 0.53$ & $5.21 \pm 16.71$ & $2.46 \pm 1.05$ & $1.40 \pm 1.03$  \\
				
				proposed$^b$  &2000& $1.64 \pm 0.43$ & $3.18 \pm 3.34$ & $\textbf{2.30} \pm \textbf{0.96}$ & $\textbf{1.24} \pm \textbf{0.89}$  \\
				
				proposed$^b$  &5000& $\textbf{1.61} \pm \textbf{0.41}$ & $4.67 \pm 16.96$ & $\textbf{2.30} \pm \textbf{0.96}$ & $1.25 \pm 0.91$  \\
				
				\hline
				
		\end{tabular}}
		\label{table:msd}
	\end{table}
	
	\begin{table}[t]
		\centering
		\setlength{\tabcolsep}{3.5pt}
		\caption[Table caption text]{ 95\% HD value of the target volumes and OARs.}
		\resizebox{0.485\textwidth}{!}{
			\begin{tabular}{c*{7}{c}c}
				&$\#$Itr&Prostate$^{\dagger\ddagger}$&SV$^{\dagger\ddagger}$&Rectum& Bladder$^{\dagger\ddagger}$ \\ \hline
				&& $\mu \pm \sigma$ & $\mu \pm \sigma$ & $\mu \pm \sigma$ & $\mu \pm \sigma$ \\ \hline
				
				base$^a$ && $9.1 \pm 3.7$ & $12.2 \pm 10.7$ & $14.4 \pm 7.0$ & $8.7 \pm 12.9$ \\ 
				
				proposed$^a$ &500& $7.9 \pm 2.8$ & $9.2 \pm 8.4$ & $14.7 \pm 6.7$ & $7.9 \pm 6.7$ \\  
				
				proposed$^a$ &2000& $6.4 \pm 2.1$ & $\textbf{8.6} \pm \textbf{9.0}$ & $14.3 \pm 6.7$ & $7.2 \pm 6.5$ \\ 
				
				proposed$^a$ &5000& $5.9 \pm 2.4$ & $8.8 \pm 10.6$ & $14.0 \pm 6.6$ & $6.5 \pm 6.1$ \\ \hline
				
				base$^b$ && $6.9 \pm 1.9$ & $15.5 \pm 9.7$ & $13.9 \pm 9.5$ & $7.0 \pm 4.5$  \\ 
				
				proposed$^b$ & 500& $6.1 \pm 1.7$ & $14.6 \pm 20.4$ & $13.5 \pm 9.1$ & $6.2 \pm 5.8$  \\ 
				
				proposed$^b$ & 2000& $\textbf{5.2} \pm \textbf{1.4}$ & $11.9 \pm 9.3$ & $13.5 \pm 8.4$ & $\textbf{5.6} \pm \textbf{5.3}$  \\ 
				
				proposed$^b$ & 5000& $\textbf{5.2} \pm \textbf{1.4}$ & $13.9 \pm 20.9$ & $\textbf{13.3} \pm \textbf{8.5}$ & $5.7 \pm 5.6$  \\ 
				
				\hline
				
		\end{tabular}}
		\label{table:hd}
	\end{table}
	
	\bibliographystyle{unsrt}
	\bibliography{references}

\begin{thebibliography}{1}

\bibitem{NCS}
National~Cancer Institute.
\newblock {Cancer Stat Facts: Prostate Cancer}.
\newblock \url{https://seer.cancer.gov/statfacts/html/prost.html}, 2019.

\bibitem{Zhang}
Baoshe Zhang, Sung-Woo Lee, et~al.
\newblock Action levels on dose and anatomic variation for adaptive radiation
  therapy using daily offline plan evaluation: Preliminary results.
\newblock {\em Practical radiation oncology}, 9(1):49--54, 2019.

\bibitem{Sonke}
Jan-Jakob Sonke, Marianne Aznar, et~al.
\newblock Adaptive radiotherapy for anatomical changes.
\newblock In {\em Seminars in radiation oncology}, volume~29, pages 245--257.
  Elsevier, 2019.

\bibitem{Yan}
Di~Yan, Frank Vicini, et~al.
\newblock Adaptive radiation therapy.
\newblock {\em Physics in Medicine \& Biology}, 42(1):123, 1997.

\bibitem{Qiao}
Qiao Yuchuan.
\newblock {\em Fast Optimization Methods For Image Registration In Adaptive
  Radiation Therapy}.
\newblock {PhD} dissertation, Leiden University Medical Center, 2017.

\bibitem{Elmahdy}
Mohamed~S. Elmahdy, Thyrza Jagt, et~al.
\newblock Robust contour propagation using deep learning and image registration
  for online adaptive proton therapy of prostate cancer.
\newblock {\em Medical physics}, 2019.

\bibitem{Milletari}
Fausto Milletari, Nassir Navab, et~al.
\newblock V-net: Fully convolutional neural networks for volumetric medical
  image segmentation.
\newblock In {\em Fourth International Conference on 3D Vision (3DV)}, pages
  565--571. IEEE, 2016.

\bibitem{Tong}
Nuo Tong, Shuiping Gou, et~al.
\newblock Fully automatic multi-organ segmentation for head and neck cancer
  radiotherapy using shape representation model constrained fully convolutional
  neural networks.
\newblock {\em Medical physics}, 45(10):4558--4567, 2018.

\bibitem{Reddi}
Sashank~J. Reddi, Satyen Kale, et~al.
\newblock On the convergence of adam and beyond.
\newblock {\em arXiv preprint arXiv:1904.09237}, 2019.

\end{thebibliography}

\end{document}